\documentclass[12pt]{article}
\usepackage[pdftex]{graphicx} 
\usepackage{epsfig}
\usepackage{comment}
\usepackage{latexsym}
\usepackage{color}
\newcommand{\mysquare}[0]{\raise-.2ex\hbox{{\Large$\Box$}}}
\def\lsim{\mathrel{\rlap {\raise.5ex\hbox{$ < $}}
{\lower.5ex\hbox{$\sim$}}}}
\def\gsim{\mathrel{\rlap {\raise.5ex\hbox{$ > $}}
{\lower.5ex\hbox{$\sim$}}}} \topmargin -1.5cm \textheight=22.5cm \textwidth=16.5cm
\catcode`\@=11
\newcount\hour
\newcount\minute
\newtoks\amorpm
\hour=\time\divide\hour by60 \minute=\time{\multiply\hour by60
\global\advance\minute by-\hour}
\edef\standardtime{{\ifnum\hour<12 \global\amorpm={am}%
        \else\global\amorpm={pm}\advance\hour by-12 \fi
        \ifnum\hour=0 \hour=12 \fi
        \number\hour:\ifnum\minute<10 0\fi\number\minute\the\amorpm}}
\edef\militarytime{\number\hour:\ifnum\minute<10 0\fi\number\minute}
\def\draftlabel#1{{\@bsphack\if@filesw {\let\thepage\relax
   \xdef\@gtempa{\write\@auxout{\string
      \newlabel{#1}{{\@currentlabel}{\thepage}}}}}\@gtempa
   \if@nobreak \ifvmode\nobreak\fi\fi\fi\@esphack}
        \gdef\@eqnlabel{#1}}
\def\@eqnlabel{}
\def\@vacuum{}
\def\draftmarginnote#1{\marginpar{\raggedright\scriptsize\tt#1}}
\def\draft{\oddsidemargin -.2truein
        \def\@oddfoot{\sl preliminary draft \hfil
        \rm\thepage\hfil\sl\today\quad\militarytime}
        \let\@evenfoot\@oddfoot \overfullrule 3pt
        \let\label=\draftlabel
        \let\marginnote=\draftmarginnote
   \def\@eqnnum{(\theequation)\rlap{\k

 ern\marginparsep\tt\@eqnlabel}%
\global\let\@eqnlabel\@vacuum}  }

\newcommand{\be}[0]{\begin{equation}}
\newcommand{\ee}[0]{\end{equation}}
\newcommand{\ba}[0]{\begin{eqnarray}}
\newcommand{\ea}[0]{\end{eqnarray}}

\def\bs{\begin{subequations}}
\def\es{\end{subequations}}

\def\thebibliography#1{%
\vskip 0.5cm \centerline{\bf \Large References}
\list{%
[\arabic{enumi}]}{\settowidth\labelwidth{[#1]} \leftmargin\labelwidth
\advance\leftmargin\labelsep
\usecounter{enumi}}
\def\newblock{\hskip .11em plus .33em minus .07em}
\sloppy\clubpenalty4000\widowpenalty4000 \sfcode`\.=1000\relax}

\renewcommand{\theequation}{\arabic{section}.\arabic{equation}}

\renewcommand{\section}{\setcounter{equation}{0}\@startsection
{section}{1}{0mm}{-\baselineskip}{0.5\baselineskip} {\normalfont\Large\bfseries}}

\renewcommand{\subsection}{\@startsection
{subsection}{2}{0mm}{-\baselineskip}{0.5\baselineskip} {\normalfont\large\bfseries}}

\renewcommand{\subsubsection}{\@startsection
{subsubsection}{3}{0mm}{-\baselineskip}{0.5\baselineskip}
{\normalfont\normalsize\slshape}}

\usepackage{amsmath}
\usepackage{amssymb,amsfonts}
\usepackage{graphicx}
\usepackage{cite}

\renewcommand{\and}{\mbox{and}}

\begin{document}
\begin{titlepage}
\begin{flushright}
LPTENS-10/07, January 2010 
\end{flushright}

\vspace{2cm}

\begin{centering}
{\bf\huge Introduction to Superstring }

{\bf\huge Cosmology}\\

\vspace{2cm}
 {\Large Costas~Kounnas  \\
 }

\vspace{1cm}

 Laboratoire de Physique Th\'eorique,
Ecole Normale Sup\'erieure,$^\dagger$ \\
24 rue Lhomond, F--75231 Paris cedex 05, France\\
\vspace{2mm}

{\em  Costas.Kounnas@lpt.ens.fr}

\
 \vspace{2cm}

{\bf\Large Abstract}

\end{centering}

\begin{quote}
This is a summary of lectures in superstring cosmology given by the author at the CORFU 2009 School and Workshops ``Theory - Cosmology - Phenomenology'',
Corfu Institute, Greece, Sept 6-13, 2009. These lectures are based on some recent developments and ideas, in the framework of superstring theory, concerning  the evolution and structure of the universe in (i) the very early ``non-geometric'' cosmological era, (ii) the intermediate ``radiation-like'' era and (iii) the late time cosmological era characterized by the electroweak phase transition.
\noindent 
\end{quote}
\vspace{5pt} \vfill \hrule width 6.7cm \vskip.1mm{\small \small \small \noindent $^\ast$\ Research
partially supported by ANR (CNRS-USAR) contract 05-BLAN-0079-02, the  IFCPAR programme 4104-2 and the ANR programme blanc
NT09-573739.\\
$^\dagger$\ Unit{\'e} mixte  du CNRS et de l'Ecole Normale Sup{\'e}rieure associ\'ee \`a
l'Universit\'e Pierre et Marie Curie (Paris
6), UMR 8549.}

\end{titlepage}
\newpage
\setcounter{footnote}{0}
\renewcommand{\thefootnote}{\arabic{footnote}}
 \setlength{\baselineskip}{.7cm} \setlength{\parskip}{.2cm}

\setcounter{section}{0}


\section{Introduction}

The next decade will witness a new era of close collaboration between theorists and experimentalists in High Energy Physics, Astrophysics and Cosmology. The LHC will provide collisions of 7 $\rm TeV$ protons, opening an entirely new high energy domain to experimental and theoretical investigation. Particles at such high energy scales interacted in the very early Universe era, fractions of a second after the Big Bang. In addition, there is by now a plethora of observational data (WMAP experiment, type Ia supernovae, the large scale distribution of matter in the Universe), favoring a phenomenological version of hot Big Bang cosmology. The existing cosmological data will be enriched by upcoming experiments, such as the launch of the European PLANCK satellite that will provide accurate measurements of the high multiple moments of the Cosmic Microwave Background (CMB) temperature fluctuations.

Both the Standard Model and the theory of General Relativity are consistent with all existing particle physics experiments and with the
astrophysical and cosmological observations. This understanding, however, is incomplete because of the following reasons:\\$\bullet$ General Relativity is a classical theory, not valid at the quantum level.\\
$\bullet$  The Standard Model involves a large number of  parameters, some of which are unnaturally small.\\$\bullet$ Neutrino oscillations and the convergence of gauge coupling constants point to the existence of a heavy mass scale, the Grand Unification scale, $M_U\sim 10^{16} $GeV, and/or the string scale $M_{\rm str} \sim 10^{17}$ Gev.\\$\bullet$ Many features of the standard cosmological model such as the very nature of dark matter and dark energy, the asymmetry between matter and antimatter, the initial Big Bang singularity (still present in inflationary models), the quantum structure of black holes and of space-time itself, are not yet well understood.

Supersymmetry, in its spontaneously broken phase, is the leading proposal for new physics beyond the Standard Model, subject to experimental tests in modern colliders, and is the leading paradigm for the early Universe and inflationary cosmology. Supersymmetry is naturally realized in the framework of superstrings, which is the only known candidate for a consistent quantum theory of gravity. Furthermore, superstring theory has successfully passed a series of nontrivial tests:\\$\bullet$  Gravity and the basic ingredients of the Standard Model emerge very naturally.\\$\bullet$  Superstring theory is free of ultraviolet divergences and anomalies.\\
$\bullet$  It incorporates in a very beautiful and natural way many of the major theoretical ideas such as 
supersymmetry, grand unification and the possible existence of large extra dimensions.\\$\bullet$ It has explained the microscopic origin of the entropy and other thermodynamical properties of black holes.

The actual understanding of superstring theory provides new important ingredients that suggest new avenues in both particle physics and cosmology, as well as in addressing some of the conceptual and technical difficulties concerning the nature of quantum gravity and cosmology. These include \cite{Superstrings}:
(i) The perturbative and non-perturbative string and M-theory dualities, which relate the strong and weak coupling phases of the theory as well as the strong and weak curvature regimes of space-time.
(ii) The discovery of objects of higher dimensionality, such as D-branes, which extend the notion of point particles and act as sources of generalized magnetic and geometrical fluxes.
(iii) The introduction of such branes and fluxes gives rise to rich families of four dimensional string vacua, 
which exhibit spontaneous supersymmetry breaking, and allow for the stabilization of some of the moduli fields at the classical, and even more importantly, at the quantum string level.
(v) The new cosmological attractor solutions in the intermediate cosmological regime, which occur before the electroweak symmetry breaking phase transition \cite{Attractor}.
(iv) Recent advances in understanding the non-geometrical structure of the initial phase of the Universe, which may provide alternative mechanisms to inflation \cite{MSDS, BKPPT}.

Despite this great list of successes however, a complete theoretical framework for studying cosmology is lacking. There are deep conceptual questions concerning the formulation of a quantum theory for cosmology at the most basic level.  
Does a wave-function description of the Universe make sense?  What are the precise cosmological observables, and how do we compute them? Extrapolating the cosmological evolution back in time, using the equations of motion of quantum field theory and classical general relativity, leads to the initial Big Bang singularity. What physics resolves such a space-like singularity? Can we continue the cosmological evolution passed the initial singularity, and finally, how do we describe in a unified theoretical framework the various phase transitions that occur throughout the evolution of the Universe? 

By studying various observational data concerning the spectrum of fluctuations of the cosmic microwave background, type Ia supernovae and the distribution of large-scale structures, we are learning more about the distribution of energy into various forms, the Cosmic pie. Only about 4$\%$ of this energy can be accounted by Standard Model particles. 22$\%$ is in the form of cold dark matter and the rest 74$\%$ is in the form of dark energy, the simplest explanation for it being a positive cosmological constant. This cosmological constant is unnaturally small, and there is no known mechanism or symmetry to explain its value. The presence of dark energy implies that the Universe itself will asymptotically approach a de Sitter universe at very late times. This implies in turn that a big portion of our universe will remain unobservable, outside the causal reach of a single observer. This late time description of the Universe still requires a deeper understanding, which is related to the very infrared behavior of gravitational and matter theories.

The best starting point to address some of these puzzles is Superstring theory, since this is the only promising candidate for a consistent theory of quantum gravity. One should investigate whether the web of string dualities can be extended to time-dependent cosmological settings. In an effort to build a concrete theoretical framework for studying cosmology, a class of string theory vacua, where the back-reaction of both thermal and quantum effects can be systematically taken into account, was recently examined in refs \cite{KTT,CosmoAll,Attractor}. 

In particular, starting with initially supersymmetric four dimensional string models, and implementing the thermal and the quantum corrections due to spontaneous breaking of supersymmetry, cosmological solutions are found, at least when the temperature $T$ and the supersymmetry breaking scale $M$ are sufficiently below the Hagedorn temperature $T_H$. In this limit the thermal and quantum stringy corrections are under control and calculable without any infrared and ultraviolet ambiguities. These corrections lead to a well-defined energy density and pressure at least in an intermediate cosmological regime, $T_H\gg T,M\ge T_W$, after the Hagedorn phase transition at $T_H$ and before the electroweak phase transition at $T_W$ .

\section{String Cosmological Phases}The finite temperature stringy setup naturally suggests a separation of the cosmological evolution in at least four distinct phases, according to the value of the temperature. Namely:\\(i) The very early phase, or even the ``(Pre-) Big Bang phase'', where the underlying string degrees of freedom are exited, or even strongly coupled. Perhaps string dualities can be applied to understand this phase and resolve the naive classical Big Bang singularity \cite{BV}.\\(ii) The stringy Hagedorn phase, $T\sim T_H$, where string oscillators and the thermal winding states must be properly taken into account. Both phases (i) and (ii) lead to a non-geometrical structure, e.g. the T-fold cosmologies studied recently in refs \cite{KTT,MSDS}. In these high temperature, high curvature and high string coupling regimes the topology and dimensionality of the space are not well-defined concepts. \\ (iii) The third phase has features similar to that of a radiation-like Freedmann cosmology. Here the Universe has cooled down to temperatures far below Hagedorn. The effects of string massive states are exponentially suppressed. The Spectator moduli, namely the moduli that are not participating in the supersymmetry breaking mechanism are either frozen at extended symmetry points, or effectively frozen by the cosmological friction, as shown recently in refs \cite{KTT,CosmoAll,Attractor} . In this phase, the ratio of the temperature T and supersymmetry breaking scale M is fixed, both evolving inversely proportional to the scale factor of the Universe. There are indications that in cases with $N$=1 initial supersymmetry, the behavior can be that of an accelerating universe at very late times.\\(iv) At lower temperatures, the effective field theory approach is valid. We are expecting new phenomena such as the electroweak phase transition, QCD confinement and structure formation to take place. We expect also that in this phase some dynamics, becoming relevant at these lower temperatures, will stabilize the no-scale modulus associated to the supersymmetry breaking scale \cite{Noscale}, realizing a cosmological, dynamical mechanism for the scale hierarchy, $M_W\ll M_{\rm Planck}$.

\subsection{\bf \it The very early (Pre-) Hagedorn Cosmological Era}In string cosmology, one must face the well-known instabilities, arising at the very early cosmological regime, where the temperature $T$ is close to the Hagedorn temperature $T_H$. At this temperature, the partition function diverges due to the exponential growth of the single particle string states as a function of mass. Many plausible scenarios concerning this very early cosmological regime have been proposed in the literature, trying to resolve or bypass the Hagedorn Cosmological Era \cite{BV}. Some of these scenarios (perturbative or even non-perturbative) include non-trivial phase transitions driven by thermal tachyon condensation, various versions of the Pre-Big Bang scenario, bypassing the initial singularity, (non)-perturbative exit scenario etc. However, all of the currently existing proposals have to overcome conceptual problems, and lack adequate quantitative control. Obviously, the deeper understanding of the Hagedorn Cosmological Era is one of most fundamental problems in string cosmology. Similar instabilities arise, not only when the temperature $T$, but also when the supersymmetry breaking scale $M$ is close to the string scale \cite{CosmoAll,Attractor}. Some interesting ideas concerning this early times, stringy cosmological phase have been presented recently in \cite{MSDS,Attractor,CosmoAll} where it was argued that the introduction of certain chemical potentials, associated with discrete gravito-magnetic fluxes, in the canonical ensemble of superstrings removes the Hagedorn instabilities. These ensembles are characterized by thermal duality for the free energy, $F(T/T_H)=F(T_H/T)$. The corresponding thermodynamics is non-standard, and the cosmological implications are under investigation. In particular, the possibility to address some of the cosmological puzzles such as the entropy, the flatness and horizon problems within this context has to be explored. If successful, this will provide a mechanism alternative to inflation.

\subsection{\bf \it The Intermediate Cosmological Era; \\the Cosmological attractor solutions} The Hagedorn transition exit is not yet well understood. The ambiguities of a plausible exit mechanism at very early cosmological times can be parameterized in terms of arbitrary initial data, by considering all possible initial conditions starting with an initial temperature $T_E$ much lower than the Hagedorn temperature $T_H$. There are similar Hagedorn-like ambiguities associated with the supersymmetry breaking scale $M$. This observation forces us to work in the regime where both $T$ and $M$ are much smaller than $T_H$. Both $T$ and $M$ evolve with cosmological time, as is dictated by the genus-one stringy quantum and thermal corrections \cite{CosmoAll}. Generically, all moduli evolve with time, including the string coupling constant, $1/g^2_{str}=S(t)$. As has been shown in \cite{Attractor}, the time behavior of the supersymmetry breaking moduli $T$ and $M$ is drastically different from all the other moduli $m_I$ that are not participating in the breaking of supersymmetry. For late cosmological times, the Universe is attracted to a Radiation-like evolution with $M(t)$ proportional to $T(t)$ and inversely proportional to the scale factor of the universe $1/a(t)$. Depending on the boundary conditions after the Hagedorn-transition exit at $T_E$, the spectator moduli $m_I$ are stabilized either at extended symmetry points, or to arbitrary points in moduli space. Thanks to this attractor mechanism, most of the Hagedorn exit ambiguities are washed out at later cosmological times. All spectator moduli, not participating in the supersymmetry breaking, are stabilized. The attractor solutions are characterized only by the ratio of the two supersymmetry breaking scales, $T / M$, which in turn is determined by the number of initial supersymmetries and the number of vector multiplets and chiral multiplets \cite{Attractor}.  

Although the attractor solution is rigorously proven in the class of string vacua where the supersymmetry breaking is induced by non-trivial geometrical fluxes, we are expecting however its validity to other supersymmetry breaking schemes involving more general fluxes. A tool to be used
for a more general proof of the attractor solutions is string-string and M-theory dualities that convert the perturbative geometrical fluxes to some non-geometrical, non-perturbative fluxes.

\subsection{ \bf \it The electroweak phase transition in the late cosmological times}
As we already mention, in the intermediate cosmological era the cosmological evolution is attracted to a radiation-like evolution \cite{CosmoAll,Attractor}. This behavior is universal and stable at late times in certain physically relevant supersymmetry breaking schemes (structure of the fluxes). This phase however is interrupted at much later times due the appearance of new infrared scales, like for instance the various infrared renormalization group invariant scales $Q$ associated with: (i) $\Lambda_G$ of hidden gauge group(s), (ii) the transmutation scale(s) $Q_H$, (induced in the infrared), by the renormalized structure of the ``soft supersymmetry breaking terms''. The existence of  a non-trivial dynamical scale $Q$ modifies the very late cosmological evolution, namely after the electroweak phase transition, around $T_W\sim Q \sim  \rm {\cal O}$(1TeV ). The infrared transmutation scale(s) $Q$ is(are) irrelevant when $M(t), T(t) \gg Q$. It becomes relevant and stops the evolution of $M(t)$ when $T(t) \sim Q$ and the electroweak phase transition takes place. The behavior in the intermediate cosmological regime suggests a natural solution to the Hierarchy problem. Indeed, extrapolating the attractor solution up to the low energy regime, where $T \sim M $= $\rm {\cal O}$(1TeV ), one finds, that the value of the supersymmetry breaking scale $M(t)$ is naturally small around the electroweak phase transition, independently of its initial value at early cosmological times at $T_E$.\

 Restricting to cases where the suprsymmetry breaking is generated via geometrical fluxes implies the existence of at least one relatively large compact dimension (the one which is associated to the supersymmetry  breaking). This is by far not in contradiction with experimental results both in particle physics and cosmology. In fact, for several years the possibility of large extra dimensions has attracted the attention of the particle physics community. The future data analysis at the LHC and elsewhere includes searches for signals indicating the existence of large extra dimensions, which is the characteristic prediction of supersymmetry breaking via geometrical fluxes. On the other hand, many other choices for supersymmetry breaking exist. Most such cases are not well adapted for precise string calculations. In all other supersymmetry breaking mechanisms, we are forced to work in the effective supergravity framework rather than the full string level. It is of main interest to generalize the stringy cosmological solutions for cases involving more general supersymmetry breaking schemes and examine in more details the electroweak symmetry breaking phase transition in the cosmological and particle physics context.

\section{ Non-geometric structure of the very early Universe; \\
Dynamical emergence of space-time} 

Very recently the existence of a new kind of massive boson/fermion degeneracy symmetry was discovered in the framework of string theory. The target space-time is two-dimensional with an enhanced, non-abelian gauge symmetry \cite{MSDS}. These exotic string vacua show  a novel Massive Spectrum Degeneracy Symmetry, $MSDS$; the massive bosonic and fermionic degrees of freedom exhibit a Spectrum Degeneracy Symmetry, whereas massless bosonic and massless fermionic states are unpaired. This property distinguishes $MSDS$ theories from ordinary supersymmetric constructions. They are constructed on a $d$=2 target space-time combined with an internal space based on a compact (highly curved) non-abelian group manifold with a characteristic curvature close to the string scale. The $MSDS$-vacua are free of tachyons and other pathological instabilities. In this respect, it is natural to consider them as the most serious candidates suitable to describe the very early ``stringy non-geometric era'' of the universe. At this early epoch, classical gravity is not valid anymore and has to be replaced by a more fundamental singularity-free theory. 

In order to support the above cosmological conjecture and make it physically relevant, it is necessary to connect the initial two dimensional  $MSDS$ vacua to higher dimensional vacua (namely $d$=4), with spontaneously broken space-time supersymmetry in late cosmological times. The existence of these connections is indeed a non-trivial task, both technically and conceptually. However, some preliminary results in this direction have been already obtained in refs \cite{MSDS} showing that (large) marginal current-current deformations of the $MSDS$ vacua connect them at least ``adiabatically'' to higher-dimensional, conventional $N$=1 chiral superstring vacua. More interesting, in the emerging higher-dimensional space-time the $N$=1 supersymmetry appears to be spontaneously broken by ``geometrical fluxes'', with a well defined thermal interpretation in some Euclidean version of the models \cite{MSDS}. Furthermore, the emerging effective field theory description of the (large-) deformed $MSDS$-vacua is well described in terms of specific gauged-supergravity theories. 

A lot of work is necessary to select the initial $MSDS$-vacuum that would lead dynamically to the precise structure of our universe in late cosmological times. All the above encouraging results strongly indicate that we were in a good direction. Furthermore, the qualitative infrared behavior of the string effective ``no-scale supergravity'' field theory, strongly suggests that we are definitely in an interesting ``non-singular string evolutionary scenario'' connecting particle physics and cosmology.

 \section*{Acknowledgement}
I would like to thank my collaborators F.~Bourliot,  T.~Catelin-Jullien, J.~Estes, I. Florakis, J.~Troost and especially  H.~Partouche and N.~Toumbas, for working and exchanging conceptual ideas with me on this fascinating topic.  It was a real  pleasure to participate in this exceptional conference. Work
partially supported by ANR contract 05-BLAN-0079-02, the  IFCPAR programme 4104-2 and the ANR programme blanc
NT09-573739.

\end{document}